\documentclass[aps,showpacs,twocolumn]{revtex4}

\newcommand{\nl}{\nonumber \\}
\newcommand{\be}{\begin{equation}}
\newcommand{\ee}{\end{equation}}
\newcommand{\bea}{\begin{eqnarray}}
\newcommand{\eea}{\end{eqnarray}}
\newcommand{\Eq}[1]{Eq.~(\ref{#1})}

\begin{document}
\author{Li-Xiang Cen and Xin-Qi Li}
\affiliation{
Department of Chemistry, Hong Kong University of Science and
Technology, Kowloon, Hong Kong, and \\
NLSM, Institute of Semiconductors, 
Chinese Academy of Sciences, P.~O.~Box 912, Beijing 100083, China}

\author{YiJing Yan}
\affiliation{Department of Chemistry, Hong Kong University of Science and
Technology, Kowloon, Hong Kong}

\title{Characterization of entanglement transformation 
via group representation theory}

\begin{abstract}
  Entanglement transformation of composite quantum systems
is investigated in the context of group representation 
theory. Representation of the direct product group
$SL(2,C)\otimes SL(2,C)$, composed of local operators 
acting on the binary composite system, is realized in 
the four-dimensional complex space in terms of a set 
of novel bases that are pseudo orthonormalized.
The two-to-one homomorphism is then established for the group 
$SL(2,C)\otimes SL(2,C)$ onto the $SO(4,C)$. It is shown that
the resulting representation theory leads to the complete
characterization for the entanglement
transformation of the binary composite system.
\end{abstract}


\maketitle

Nonlocal correlation or entanglement between spatially 
separated quantum systems is one of the most striking 
features of quantum mechanics. In recent years, extensive 
attention has been addressed to this topic due to its
great importance in quantum information 
processing \cite{ekert,bennett,steane}. One basic 
approach to this subject, first exploited by Bennett and 
co-workers \cite{bennett1,bennett2}, is to employ entanglement 
transformation for quantum states under actions
of local quantum operations and classical communication (LOCC). 
Such studies, contributed to characterize what 
is possible locally and what is impossible, 
promise in principle 
criterion not only for distinguishing entangled 
(nonseparable) states from separable states but also for 
classifying entanglement in different categories. Entanglement
transformation also relates to the measures of
entanglement. For instance, the asymptotic measures
such as distillable entanglement and 
entanglement cost owe their very existence to this
body of knowledge \cite{horodecki,donald}.

Despite it is very fruitful for pure entangled states 
\cite{nielsen,vidal1,vidal2}, the entanglement 
transformation has so far provided only fragments of the 
knowledge on mixed entangled states 
\cite{linden,kent,jonathan}, even for the simplest binary 
composite system. Consider a quantum system 
described by a density matrix $\rho $ in the Hilbert space 
$H_{A}\otimes H_{B}$. State transformation under the
actions of LOCC can be illustrated as \cite{benjamin}
\be
\rho \rightarrow \rho ^{\prime }=\frac{\sum_{i,j}A_{i}
\otimes B_{j}\rho A_{i}^{\dagger }\otimes B_{j}^{\dagger }}
{\sum_{i,j}{\rm tr}(A_{i}^{\dagger}A_{i}\otimes B_{j}^{\dagger }
B_{j}\rho )},  
\label{process}
\ee
where $A_{i}$ and $B_{j}$ are local operators acting on the 
two isolated subsystems, obeying the conditions 
$\sum_{i}A_{i}^{\dagger}A_{i}\leq I$ and 
$\sum_{j}B_{j}^{\dagger }B_{j}\leq I$, respectively.
Note that, the entangled state $\rho $ by definition
does not allow local description; that is, it cannot 
be written as statistical mixtures of product states, 
$\rho \neq \sum_{i}p_{i}\rho _{A}^{i}\otimes \rho _{B}^{i}$. 
The unfavorable contrast  between the state nonlocality 
and the $A_{i}\otimes B_{j}$ action locality causes the 
characterization of the state transformation 
Eq.~(\ref{process}) extremely ambiguous. 
A different aspect to account for this problem is that the 
representation of the mixed state in terms of pure states 
is not unique, making the characterization for the nonlocal
parameters (or entanglement monotones \cite{monotone}) 
notoriously difficult.

Bearing in mind that a proper uniform description for 
both the states and operators is impossible in terms 
of local bases, we shall ask 
instead %666, on the contrary, 
the following question: is there a profitable way to describe the 
local operators $A_{i}\otimes B_{j}$ in terms of 
{\it nonlocal bases} so that the concerning problem could become more 
transparent? This question will be specified clearly
in the paper by the underlying language of group 
representation theory. 
%666Particularly, 
Note that each 
local operator $A_{i}$ ($B_{j}$) acting on the 
$N$-dimensional complex space corresponds, 
up to a real coefficient, to an
element of the unimodular linear transformation group 
$SL(N,C)$. Thus, the problem stated above can be 
recast %666outlined 
as finding out a proper realization, in terms of 
nonlocal bases, of the homomorphic representation for 
the direct product group $SL(N,C)\otimes SL(N,C)$.

In this paper we establish such a realization for the group
$SL(2,C)\otimes SL(2,C)$ that completely resolves the entanglement 
transformation of the binary composite system. 
We construct, based on 
the ``spin flip'' operation introduced by Wootters \cite{wootters}, 
a class of pseudo-orthonormal bases in the four-dimensional
Hilbert space. 
We then prove %666It is shown 
that the representative matrices 
of the group elements of $SL(2,C)\otimes SL(2,C)$ embedded 
in such pseudo-orthonormal bases are complex orthogonal, and 
thus establish %We thus set up 
the homomorphism of $SL(2,C)\otimes SL(2,C)$ 
onto the group $SO(4,C)$. 
Finally, %Subsequently, 
we show that the 
%666established 
realization of the representation for 
$SL(2,C)\otimes SL(2,C)$ actually enables the complete 
characterization for the entanglement transformation of 
the binary composite systems under LOCC.

Let us start by introducing the following notation. 
For the states in the space $C^{2}\otimes C^{2}$, define 
the {\it pseudo inner product} as 
\begin{equation}
C(\psi _{1},\psi _{2})\equiv \langle \psi _{1}|\tilde{\psi}_{2}\rangle .
\label{pproduct}
\end{equation}
Here the spin-flipped state $|\tilde{\psi}\rangle $ 
is introduced according to 
Wootters' definition \cite{wootters} 
\begin{equation}
|\tilde{\psi}\rangle =\sigma _{2}\otimes \sigma _{2}
|\psi ^{\ast }\rangle ,
\label{tilde}
\end{equation}
with $|\psi ^{\ast }\rangle $ indicating the complex 
conjugate of $|\psi \rangle $. A state $|\psi \rangle $ 
is said to be {\it pseudo-normalized} if there exists 
\begin{equation}
C(\psi )=\langle \psi |\tilde{\psi}\rangle =1.  \label{pnorm}
\end{equation}
For an arbitrary state $|\phi \rangle $ in $C^{2}\otimes 
C^{2}$, its pseudo normalization (denoted as $||\cdots ||$ 
hereafter) can be given as $|||\phi \rangle ||=re^{i\theta }
|\phi \rangle $, where $r$ is a positive real number
and the phase factor $\theta $ is set 
as $0\leq \theta <\pi $. 
Note that the pseudo normalization (\ref{pnorm}) 
avoids %666excludes 
the phase uncertainty
which occurs in the common definition of normalization.

  The set of {\it pseudo-orthonormal} bases 
$\{|x_{i} \rangle, i,j=1,\cdots ,4\}$ is defined similarly 
as the conventionally used orthonormal set but with
\begin{equation}
C(x_{i},x_{j})=\langle x_{i}|\tilde{x}_{j}\rangle =\delta _{ij}.
\label{biortho}
\end{equation}
Obviously, the pseudo-conjugate set 
\{$|\tilde{x}_{i}\rangle $\} itself constitutes also a set of
pseudo-orthonormal bases. The pseudo orthogonality 
relation (\ref{biortho}) can be alternatively stated
as the set of bases that are biorthogonal to their 
pseudo-conjugate set. From the 
biorthogonality of the dual bases it suffices to see that the 
pseudo-orthonormal basis constitutes a complete set 
in the space $ C^{2}\otimes C^{2}$. Actually, given 
an arbitrary set of linearly independent vectors 
\{$|\alpha _{i}\rangle ,i=1,\cdots ,4$\}, one can
construct immediately the pseudo-orthonormal basis 
\{$|x_{i}\rangle,i=1,\cdots ,4$\} via the following 
pseudo ``Schmidt orthogonalisation''
procedure by setting that
\bea
|x_{1}\rangle&=&|||\alpha _{1}\rangle ||,
\nl
|x_{n}\rangle &=&||(I-\sum_{i=1}^{n-1}|x_{i}\rangle 
\langle \tilde{x}_{i}|)|\alpha _{n}\rangle ||,\ \ n=2,3,4.  
\label{construct}
\eea

Before proceeding, let us mention that the special
set of maximally entangled states, the so-called 
``magic basis'' introduced by 
Bennett et al \cite{bennett2}, 
\bea
|e_{1}\rangle &=&\frac{\sqrt{2}}{2}(|00\rangle +
|11\rangle ),~~|e_{2}\rangle =i
\frac{\sqrt{2}}{2}(|00\rangle -|11\rangle ), 
\nl
|e_{3}\rangle &=&i\frac{\sqrt{2}}{2}(|01\rangle +
|10\rangle ),~~|e_{4}\rangle =\frac{\sqrt{2}}{2}
(|01\rangle -|10\rangle ),  
\label{magic}
\eea
are just the simplest pseudo-orthonormal basis. In 
this case the dual bases are completely overlapped: 
$|e_{i}\rangle =|\tilde{e}_{i}\rangle $, hence
the pseudo-orthonormal set \{$|e_{i}\rangle $\} recovers 
the common orthonormal basis. Remarkably, it is shown 
that this magic basis possesses very curious properties and 
has been employed to quantify the entanglement of some two-qubit 
states in earlier studies \cite{bennett2,hill}. Note that the 
unimodular unitary transformations acting 
locally on the two subsystems,
\begin{eqnarray}
U_{A}\otimes U_{B}&=&(\cos \theta _{a}I_{2}+
i\sin \theta _{a}\vec{r}_{a}\cdot\sigma _{a})
\nl
&&\otimes (\cos \theta _{b}I_{2}+i\sin \theta _{b}
\vec{r}_{b}\cdot\sigma _{b}),  \label{ul}
\end{eqnarray}
correspond to real orthogonal matrices in terms of
the above set of bases (\ref{magic}). In the context of
group theory, it is well known the homomorphism between
the groups $SU(2)\otimes SU(2)$ and $SO(4)$. The magic 
basis establishes in fact an explicit realization for such 
a homomorphism, namely, the two-to-one
mapping of $SU(2)\otimes SU(2)$ onto $SO(4)$ in 
the four-dimensional complex space. The kernel of the 
homomorphism is just the center of the group 
$SU(2)\otimes SU(2)$ consisting of the elements 
\{$I_{2}\otimes I_{2},-I_{2}\otimes -I_{2}$\}. In this stage, 
there is no curiosity that any real orthogonal combinations
of the magic basis correspond to local unitary transformations 
on the subsystems and hence yield still maximally entangled 
states (another magic basis).

Let us now explore the implication of the aforementioned 
pseudo-orthonormal basis $\{|x_j\rangle\}$ with \Eq{biortho},
which is obviously a nontrivial extension of the
magic basis.
To this end, consider the class of 
transformations that preserve the pseudo inner product, 
\begin{equation}
C(Rx,Ry)=C(x,y).  \label{conserv1}
\end{equation}
It is equivalent to say, by definition (\ref{pproduct}), 
that the operators $R$ satisfy 
\begin{equation}
R^{\dagger }\tilde{R}=\tilde{R}R^{\dagger }=I.  \label{conserv2}
\end{equation}
\newtheorem{theorem}{Theorem}
\begin{theorem} \label{theo1}
The statements below are equivalent:
\\ \noindent
(i) Operator $R$ preserves the pseudo inner product;
\\ \noindent
(ii) Operator $R$ transforms a set of pseudo-orthonormal bases 
\{$|x_{i}\rangle $\} into another such set 
\{$|x_{i}^{\prime }\rangle $\};
\\ \noindent
(iii) The representation of an operator $R\rightarrow D(R)$ 
in terms of the biorthogonal dual bases \{$|x_{i}\rangle $\} 
and \{$|\tilde{x}_{j}\rangle $\}, defined as $D_{ij}(R)
\equiv \langle \tilde{x}_{i}|R|x_{j}\rangle $, is a 
$4\times 4$ complex orthogonal matrix. Namely, the 
matrix $D(R)$ satisfies 
\begin{equation}
D(R)D^{T}(R)=D^{T}(R)D(R)=I,  \label{orthog}
\end{equation}
where the superscript ``$T$" denotes the matrix transpose.
\end{theorem}
{\it Proof}:
The equivalence between the statements (i) and (ii) is 
evident, merely regarding that the set of 
pseudo-orthonormal states constitute a complete set of 
bases for the space $C^{2}\otimes C^{2}$. The equivalence 
between (i) and (iii) is proven as follows. Firstly, 
given the transformation $R$ preserving the pseudo 
inner product, one has the representative matrix
\bea
(D(R)D^{T}(R))_{ij}&=&\sum_{k=1}^{4}
\langle \tilde{x}_{i}|R|x_{k}\rangle \langle 
\tilde{x}_{j}|R|x_{k}\rangle 
\nl
&=&\sum_{k=1}^{4}\langle \tilde{x}_{i}|R|x_{k}\rangle 
\langle \tilde{x}_{k}|\tilde{R}^{\dagger }|x_{j}\rangle 
\nl
&=&\langle \tilde{x}_{i}|R\tilde{R}^{\dagger }|x_{j}
\rangle =\delta _{ij},
\label{proof1}
\eea
where we have used the fact that $\sum_{k=1}^{4}|x_{k}
\rangle \langle \tilde{x}_{k}|\equiv I$ and 
$R\tilde{R}^{\dagger }=(\tilde{R}R^{\dagger })^{\dagger}=I$. 
Reciprocally, from the statement (iii) and
noting that the operator $R$ can be expressed as
$R=\sum_{ij}D_{ij}(R)
|x_{i}\rangle\langle \tilde{x}_{j}|$ 
one can obtain immediately that
\bea
R^{\dagger }\tilde{R}&=&\sum_{i,j,k=1}^{4}D_{ij}(R)
D_{ik}(R)|\tilde{x}_{j}\rangle \langle x_{k}| 
\nl
&=&\sum_{j=1}^{4}|\tilde{x}_{j}\rangle \langle x_{j}|=I.  
\label{proof2}
\eea
We have thus completed the proof.

Theorem \ref{theo1} is important as it discloses the intrinsic 
connection between the complex orthogonal matrix and the 
linear transformation that preserves pseudo inner product. 
As one can see, the aggregate of all complex orthogonal
matrices constitutes the linear transformation group 
$SO(4,C)$, which is a direct extension of the 
real orthogonal group $SO(4)$ in the magic basis. 
Physically, the presence of pseudo-orthonormal basis 
renders actually a homomorphic realization for the direct 
product group $SL(2,C)\otimes SL(2,C)$ onto $SO(4,C)$. 
\begin{theorem} \label{theo2}
The homomorphism between the transformation group
$SL(2,C)\otimes SL(2,C)$ and the complex $SO(4,C)$ can be 
realized via the pseudo-orthonormal basis.
\end{theorem}
{\it Proof}:
   Let us start with a readily verified fact that
every element of $SL(2,C)$, which is a 
complex $2\times 2$ matrix with the unit 
determinant, satisfies the relation 
\begin{equation}
A^{\dagger }\tilde{A}=I_2,\ \ B^{\dagger }\tilde{B}=I_{2}.  
\label{sl}
\end{equation}
%{\bf LX: should $I$ be $I_2$?}
Thus the operator of the direct product $A\otimes B$ 
satisfies Eq.~(\ref{conserv2}) and, hence, preserves the 
pseudo inner product. According to Theorem 1, 
the above arguments amount to the realization,
via the pseudo-orthogonal basis,
of the correspondence 
for each element $A\otimes B$ of 
the group $SL(2,C)\otimes SL(2,C)$ to a
unique complex orthogonal matrix $D$. 

  The proof of the reverse relation that 
for each matrix $D$ of $SO(4,C)$ there correspondences  
to the element $A\otimes B$ of the group $SL(2,C)\otimes SL(2,C)$
is less straightforward. 
Note that according to Theorem 1 the complex orthogonal matrix $D$ induces 
the transformation $R$ that preserves the pseudo inner 
product. Therefore, to prove the existence of $SO(4,C) \rightarrow 
SL(2,C)\otimes SL(2,C)$ correspondence, it requires only  
to demonstrate that the operator $R$ can be 
decomposed into the product form $A\otimes B$. To this 
end, let us introduce:
\\ \noindent
{\bf Lemma 1} {\it  
Any operator $H$ that transforms a set of 
pseudo-orthonormal bases \{$|x_{i}\rangle $\} into 
its dual set \{$|\tilde{x}_{i}\rangle $\} can be 
decomposed into a product form.
Such an operator $H$ is hermitian and can be written 
as $H=\sum_{i=1}^{4}| \tilde{x}_{i}\rangle \langle 
\tilde{x}_{i}|$. 
}
\\ \noindent
This Lemma was proved previously in Ref.~\cite{cen} 
through decomposing directly
the operator into product form. Note the concerning 
operator $R$ now takes the form $R=\sum_{i=1}^{4}
|x_{i}^{\prime }\rangle \langle \tilde{x}_{i}|$, which is 
generally not hermitian. It would be helpful to introduce 
the polar decomposition for the operator, 
$R=\sqrt{RR^{\dagger }}U$, where $U$ is a unitary operator. 
Then, the required property for $R$ can be proved by showing
that both the hermitian part 
$\sqrt{RR^{\dagger }}$ and the unitary $U$ can be
decomposed into the product forms.
For the hermitian part, we have explicitly that
\begin{equation}
RR^{\dagger }=\sum_{i,j=1}^{4}|x_{i}^{\prime }
\rangle \langle \tilde{x}_{i}|\tilde{x}_{j}\rangle 
\langle x_{j}^{\prime }|.  \label{part1}
\end{equation}
Suppose the pseudo-orthonormal basis \{$|\tilde{x}_{i}
\rangle $\} is related to the magic basis 
\{$|e_{i}\rangle $\} by a transformation
\begin{equation}
|\tilde{x}_{i}\rangle =\sum_{j=1}^{4}S_{ij}|e_{j}\rangle ,  
\label{trans}
\end{equation}
where $S^{T }S=I$ by Theorem 1. We 
can obtain then
\bea
RR^{\dagger }&=&\sum_{i,j,m,n}S_{im}^{\ast }S_{jn}
|x_{i}^{\prime }\rangle \langle e_{m}|e_{n}\rangle 
\langle x_{j}^{\prime }| 
\nl
&=&\sum_{m}\Bigl(\sum_{i}S_{im}^{\ast }|x_{i}^{\prime }\rangle\Bigr)
\Bigl(\sum_{j}S_{jm}\langle x_{j}^{\prime }|\Bigr)
\nl
&=&\sum_{m=1}^{4}
|y_{m}\rangle
\langle y_{m}|.  
\label{hermi}
\eea
Here $|y_{m}\rangle \equiv \sum_{i}S_{im}^{\ast }
|x_{i}^{\prime }\rangle $, which obviously constitutes 
another set of pseudo-orthonormal bases satisfying 
$\langle y_{m}|\tilde{y}_{n}\rangle =\delta _{mn}$. 
Consequently, according to Lemma 1, the operator 
$RR^{\dagger }$ (hence its square root 
$\sqrt{RR^{\dagger }}$) can be decomposed into the product 
form. 
%%%
To show the unitary part $U$ of the
operator $R$ is also factorizable, write out 
\begin{equation}
I=R^{\dagger }\tilde{R}=U^{\dagger }\bigl(\sqrt{RR^{\dagger }}\bigr)
\bigl(\widetilde{\sqrt{RR^{\dagger }}}\bigr)\tilde{U}=U^{\dagger }
\tilde{U} , 
\label{unitary}
\end{equation}
which leads to $U=\tilde{U}$. Therefore, $U$ must 
transform magic basis into another set of magic basis. 
As mentioned earlier, $U$ just belongs to the class 
of unitary transformations that act locally on the 
subsystems: $U=U_{A}\otimes U_{B}$. Note that
\begin{equation}
R^{\dagger }R=U^{\dagger }(RR^{\dagger })U.  
\label{polar}
\end{equation}
Therefore, once the product form of $R^{\dagger }R$ and 
$RR^{\dagger }$ are obtained, the distinct forms of 
$U_{A}$ and $U_{B}$ can be worked out
consequently. 
We have now concluded the required property
that the operator $R$ preserving pseudo inner product does
have a product form
$A \otimes B$.

  The above proof depicts also that the 
transformation of $SO(4,C)$ under the pseudo-orthonormal 
basis does correspond to the 
physical operation acting on the two 
subsystems locally. Obviously, the mapping
$R = A\otimes B \rightarrow D(R)$ preserves the group multiplication.
We have thus completed the proof  of Theorem 2, the
homomorphism of $SL(2,C)\otimes SL(2,C) \sim SO(4,C)$.
More specifically, by viewing the unimodular constraint of
$\det A=\det B=1$, the mapping 
between the elements of the group $SL(2,C)\otimes SL(2,C) $ 
and that of $SO(4,C)$ is two-to-one, reflected by the
kernel \{$I_{2}\otimes I_{2},-I_{2}\otimes -I_{2}$\}.

  With the group reprsentation theory established,
we are now in the position to investigate the entanglement 
transformation for the binary composite system. 
In practice, the general LOCC process (\ref{process}) 
can be decomposed into a sequence of the elementary 
actions of generalized measurements with two outcomes. 
Commonly, as in the filtering process \cite {gisin}, 
only one outcome will produce the desired result with 
a finite probability. We thus consider simply 
the following transformation: 
\begin{equation}
\sigma =\frac{A\otimes B\rho A^{\dagger }\otimes 
B^{\dagger }}{{\rm tr}(\rho A^{\dagger }A\otimes B^{\dagger }B)}. 
\label{process2}
\end{equation}
Suppose the matrix representation of the state $\rho $ 
in a set of pseudo-orthonormal bases \{$|\tilde{x}_{i}
\rangle $\} is given by $\langle \tilde{x}_{i}|\rho |
\tilde{x}_{j}\rangle =h_{ij}^{\rho }$, or explicitly,
\begin{equation}
\rho =\sum_{i,j=1}^{4}h_{ij}^{\rho }|x_{i}\rangle 
\langle x_{j}|.
\label{express}
\end{equation}
Since the operator $A\otimes B$ just transforms the 
basis \{$|x_{i}\rangle $\} into another pseudo-orthonormal 
basis, say \{$|y_{i}\rangle $\}, one has 
\begin{eqnarray}
\sigma&=&N\sum_{i,j=1}^{4}h_{ij}^{\rho }|y_{i}
\rangle \langle y_{j}|
\nl
&=&N\sum_{i,j=1}^{4}(S^{T }
h^{\rho }S^{\ast })_{ij}|x_{i}\rangle \langle x_{j}|,  
\label{trans2}
\end{eqnarray}
where $N$ is the trace norm factor and $S^{T }S=I$. 
Thus, we can now conclude the following fact:
\begin{theorem} \label{theo3}
The state transformation $\rho \rightarrow 
\sigma $ for the composite binary system can be 
realized under LOCC iff there exists a complex 
orthogonal matrix $S$ such that
\begin{equation}
\frac{h^{\sigma }}{\det h^{\sigma }}=\frac{Sh^{\rho }
S^{\dagger }}{\det h^{\rho }},  
\label{condi}
\end{equation}
where $h^{\rho }$ and $h^{\sigma }$ are representative 
matrices of $\rho $ and $\sigma $ in the pseudo-orthonormal 
basis, respectively.
\end{theorem}
Moreover, note that the density matrix $\rho $ can 
always be ``diagonalized'' in a certain 
set of pseudo-orthonormal bases \cite{wootters}. That is that
\begin{equation}
\rho =\sum_{i=1}^{4}\lambda _{i}^{\rho }|x_{i}^{\rho }
\rangle \langle x_{i}^{\rho }|,  
\label{diag}
\end{equation}
where $\lambda _{i}^{\rho }$ in the decreasing order 
has been shown to be related to the volume of a certain measure of 
entanglement in $\rho $. According to Theorem \ref{theo3},
we have diag$\{\lambda _{1}^{\rho },\lambda _{2}^{\rho },
\lambda _{3}^{\rho },\lambda _{4}^{\rho }\}$~
$\propto$~diag$\{\lambda_{1}^{\sigma },\lambda _{2}^
{\sigma },\lambda _{3}^{\sigma },\lambda _{4}^{\sigma }\}$.
Thus it can be concluded concisely that the 
transformation $\rho \rightarrow \sigma
=\sum_{i=1}^{4}\lambda _{i}^{\sigma }|x_{i}^{\sigma }
\rangle \langle x_{i}^{\sigma }|$ can be achieved by 
LOCC iff $\lambda _{i}^{\rho }/\lambda_{j}^{\rho }=
\lambda _{i}^{\sigma }/\lambda _{j}^{\sigma }$ 
\cite{linden2}. The detailed
local actions can be acquired straightforwardly as
\be \label{action}
\frac{A\otimes B}{(\det A\times \det B)^{1/2}}=\sum_{i=1}^{4}
|x_{i}^{\sigma }\rangle 
    \langle \tilde{x}_{i}^{\rho}|.  
\ee

   In summary, this work provides a fruitful application
of the group representation theory to the study of the LOCC
entanglement transformation. We have shown, for the simplest 
binary composite system, the local operators acting on the 
subsystems are represented by complex orthogonal matrices 
in terms of the set of bases that are pseudo orthonormalized. 
This leads explicitly to the notable realization of the 
homomorphic representation for the direct product group 
$SL(2,C)\otimes SL(2,C)$ onto the group $SO(4,C)$. 
Consequently, the entanglement transformation for the binary 
composite system is completely characterized by virtue of the 
resulting representation theory. Further studies on the entanglement 
transformation of more complex composite quantum systems,
such as the many-pair binary systems and the multipartite 
composite systems, should resort to the representation theory
of Lie group with high dimensions, which will be a subject in
the future research.

  This work is financially supported 
by the Postdoctoral Fellowship from HKUST,
the Postdoctoral Science Foundation of China, the
special funds for Major State Basic 
Research Project No.~G001CB3095 of
China, and the special grant from Chinese Academy of
Sciences to distinguished young researchers.


\begin{thebibliography}{99}
\bibitem{linden2} It was first proven in Ref.~\cite{linden} 
that the ratios $\lambda _{i}/\lambda_{j}$ are invariant under
the LOCC operations.
\end{thebibliography}
\end{document}